\documentclass[multphys,vecphys]{svmult}

\usepackage{makeidx}         % allows index generation
\usepackage{graphicx}        % standard LaTeX graphics tool
\usepackage{multicol}        % used for the two-column index
\usepackage[bottom]{footmisc}% places footnotes at page bottom
\makeindex             % used for the subject index
\begin{document}

\title*{Modeling of the Terminal Velocities of the Dust Ejected Material by the Impact}
\titlerunning{Modeling of the Terminal Velocities}
%for an abbreviated version of
\author{M. Rengel\inst{1},
M. K\"uppers\inst{1}, H.U. Keller\inst{1}, and P. Gutierrez\inst{2}}
\institute{Max-Plack-Institut f\"ur Sonnensystemforschung,
Max-Planck-Strasse 2, 37191,
    Katlenburg-Lindau, Germany.
\texttt{rengel@mps.mpg.de} \and Instituto de Astrof\'isica de
Andalucia-CSIC.C/Camino Bajo de Hu\'etor, 50,
    18008, Granada, Spain.
%    \texttt{name@email.address}
}
\maketitle

\section{Abstract}
\label{abstract} We compute the distribution of velocities of the
particles ejected by the impact of the projectile released from NASA
Deep Impact spacecraft  on the nucleus of comet 9P/Tempel 1 on the
successive 20 hours following the collision. This is performed by
the development and use of an ill-conditioned inverse problem
approach, whose main ingredients are a set of observations taken by
the Narrow Angle Camera (NAC) of OSIRIS onboard the Rosetta
spacecraft, and a set of simple models of the expansion of the dust
ejecta plume for different velocities. Terminal velocities are
derived using a
maximum likelihood estimator. \\

We compare our results with published estimates of the expansion
velocity of the dust cloud. Our approach and models reproduce well
the velocity distribution of the ejected particles. We consider
these successful comparisons of the velocities as an evidence for
the appropriateness of the approach. This analysis provides a more
thorough understanding of the properties of the Deep Impact dust
cloud.

\section{Introduction}
\label{sec:1} In order to investigate the comet's interior, on 4
July 2004 the NASA mission Deep Impact fired a projectile of 364 kg
into the nucleus of 9P/Tempel 1. A crater was formed and material
was ejected from the comet~\cite{a05:ref}. OSIRIS, the scientific
camera system on the Rosetta spacecraft~\cite{k05:ref}, was
activated five days before the Deep Impact event, and observed
9P/Tempel 1 near-continuously for more than two weeks
\cite{2005Natur.437..987K:ref}. Some properties of the cloud (e.g.
structure, morphology, number of water molecules produced in the
impact, abundance ratio between the CN parent molecules and water)
are already reported in the literature.

The main aim of this work is to present the velocity distribution of
the ejected dust computed with a discrete linear inverse approach.
The Narrow Angle Camera (NAC) of OSIRIS monitored the cometary dust
of 9P/Tempel 1, and by aperture photometry on the images, light
curves for different circular fields provide the observational input
to our approach. A model of the expansion of the dust ejecta plume
for different velocities supplies the modeled light curves. Because
of the good temporal parameter sampling that the OSIRIS data offer
(into around an image per minute), this approach is very well
suitable. The comparison between these results and those derived
from other authors not only tests and refines the conclusions
derived from other observations and methods, but also provides an
evidence for the appropriateness of our models and of our approach,
and furthermore, allow to test of additional production of the
material.

% Always give a unique label
% and use \ref{<label>} for cross-references
% and \cite{<label>} for bibliographic references
% use \sectionmark{}
% to alter or adjust the section heading in the running head
%Your text goes here. Use the \LaTeX\ automatism for your citations
%\cite{monograph}.

\section{Observational data and models} \label{sec:2}

The observational data consist of 73 images of 9P/Tempel 1 taken by
the NAC OSIRIS with the clear filter, from the impact time to around
20 hours after impact time (for a detailed description of the OSIRIS
observations, see \cite{k06:ref}). By cometocentric circular
aperture photometry on the images of 9P/Tempel 1, the brightness of
the comet is integrated over a set of nine apertures of different
radii (from 3000 to 15000 km, 1 pixel = 1500 km). An observational
data row matrix $O$ is defined, and contains the collection of
integrated brightness (73 images $\times$ 9 apertures = 657
data-points):
\\*

$O$ = [O$_{1}$, O$_{2}$,...,O$_{73}$,...,O$_{657}$].
\\*

The measurement error row matrix E of the observed brightness (or
the known individual standard deviation matrix) is defined by:
\\*

$E$ = [$\sigma_{1}$,
$\sigma$$_{2}$,...,$\sigma$$_{73}$,...,$\sigma$$_{657}$].
\\

The models of the expansion of the dust ejecta plume are computed
using the decay of the flux when the material leaves the different
apertures. For early times (less than 20 hours), we assume: (1) that
the expanding dust cloud moves with a constant velocity, (2) a
neglected effect of the radiation pressure removing the material,
(3) and that the dust reaches its final velocity instantly at the
time of the impact. We compute the fluxes of the dust ejecta plume
for different projected ejecta velocities $v$, from 1 to 600
ms$^{-1}$, with a steep $i$ of 50 ms$^{-1}$. We adopt a maximum
velocity of 600 ms$^{-1}$, although a small fraction of the dust
might be actually be ejected at higher velocities ~\cite{a05:ref}. A
conservative minimum velocity of 1 ms$^{-1}$ is adopted. The models
are normalized in flux by multiplying each modeled flux by a
normalization constant. The generated normalized models are stored
in a matrix $M$:

\[ M = \left[ \begin{array}{cccc}
M_{1,1}  & M_{1,2}  & ... & M_{1,657}\\
M_{2,1}  & M_{2,2}  & ... & M_{2,657} \\
& & & \\
M_{13,1} & M_{13,2} & ... & M_{13,657} \end{array} \right]\]
\\*
where the first sub-index $j$ of each element denotes the velocity
bin $i$ and the second sub-index $k$ the modeled brightness point.
As an illustration, Fig.1 shows the lightcurves of the cometary dust
for nine circular fields from 2 to 10 pixels, and models with
projected velocities of 110, 160 and 300 m s$^{-1}$.

\begin{figure}
\begin{center}

\includegraphics[angle=-90,width=.9\textwidth]{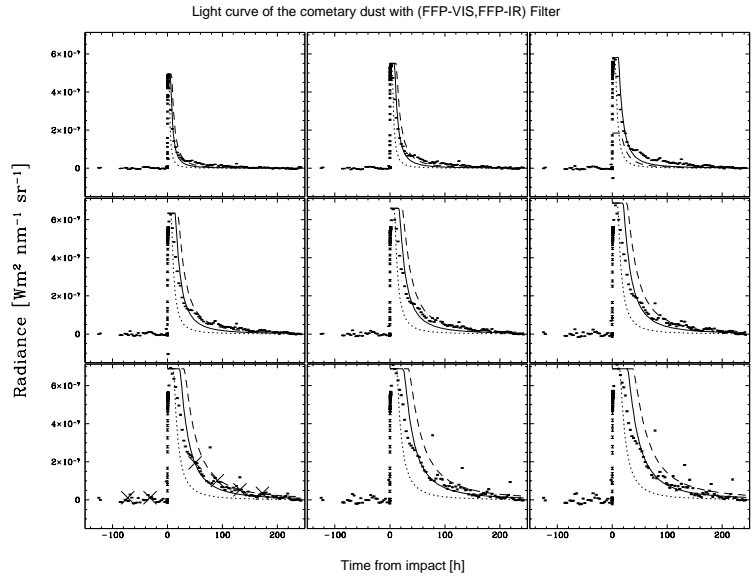}
\\
\end{center}
\caption{ Light curves of the cometary dust in the clear filter, for
nine circular fields from 2 to 10 pixels (or from 3000 to 15000 km
radius at the comet) from left to right. The dashed, solid and
dotted lines show model profiles for velocities of 110, 160, and 300
m/s, respectively. The crosses on the bottom left panel are
separated by 40.832 h, the rotation period of comet 9P/Tempel 1.}
\end{figure}

\section{Approach} \label{sec:2}

We now consider an approach to retrieve the velocity distribution of
particles. Instead to simultaneously fit synthetic images to a
selected set of observations (e.g. \cite{jorda06:ref}), we
simultaneously fit synthetic light-curves to the observed ones. In
the earlier method, in order to clean the images from eventual
residuals of cosmic rays and to improve the S/N ratio, new images
should be created by stacking a set of successive images. This would
decrease the temporal parameter sampling. Because we are extracting
the brightness from each image, our method keeps the good OSIRIS
temporal parameter sampling.
%A velocity distribution computed from a good temporal parameter
%sampling instead from a defines a distribution with a bigger
%temporal resolution than a spatial one.
%, which adjudicates a compromise between two.
Given the matrices $O$, $E$, and $M$, our focus is to find the curve
or the most likely model, among all combination of simple models,
that approximately fits the sets of observations, i.e., the observed
brightness distribution of the cometary dust. Distribution of
terminal velocities reached by the dust are derived using a maximum
likelihood estimator and a least squares solution.

We fit simultaneously the observed fluxes O by minimizing the sum of
the chi-squared differences between all the data-points $O$ and
linear combinations of the modeled fluxes $M$:

\begin{equation}
\chi^2=\sum_{k=1}^{657}  \frac{1}{\sigma_{k}^2} \left(
\sum_{j=1}^{13} [a(j)]_k - O_k \right)^2
\end{equation}
\\*

where $[a(j)]_k$ is defined as

\begin{equation}
[a(j)]_k=c_jM_{jk}
\end{equation}
\\*

and $c$, a set of positive dimensionless coefficients to be
determined, as:

\begin{equation}
c=(c_1,c_2,...,c_{13}).
\end{equation}

The minimization of the expression (1) occurs where the derivate
with respect to all parameters $c_j$ vanishes. The so-called normal
equations of the least-squares problem can be expressed as:

\begin{equation}
\frac{\partial \chi^2}{\partial c_j}=2\sum_{k=1}^{657}
\frac{1}{\sigma_{k}^2}\left( \sum_{j=1}^{13} c_jM_{jk} - O_k
%\right)M_{jk}\delta_{k}^{j}=0
\right)M_{jk}=0
\end{equation}
\\*

%where
%
%\begin{equation}
%\delta_{k}^{j}=\frac{\partial c_j}{\partial c_k}
%\end{equation}

Eq. 4 yields a linear system of 13 equations which are solved using
the LAPACK library ~\cite{LAPACK3:ref}. The solution corresponds to
the deepest minimum of the chi-square function (1). There may be
other solutions besides it which corresponds to local minima of the
function (1) (with a similar likelihood). This means that our
derived parameters $c$ may be affected by large errors.

\section{Results}

In Fig. 2 we show the velocity distribution . The dust particles
have velocities of $\sim$225 ms$^{-1}$. This velocity is in a good
agreement with the projected speed values derived from other authors
(e.g. ~\cite{meech05:ref} (200 ms$^{-1}$), ~\cite{sugita05:ref} (125
ms$^{-1}$), ~\cite{schleicher06:ref} (130-230 ms$^{-1}$),
~\cite{jorda06:ref} (190 ms$^{-1}$)). Nevertheless,
\cite{feldman06:ref}, from HST images, derive lower velocities
(70-80 ms$^{-1}$).

Dust with a velocity of 200 ms$^{-1}$ reaches the edge of the
smallest aperture  (2-pixel radius) in slightly more than 4 h. The
acceleration phase is not seen, meaning that not more than a few
percent of the impact material was produced later than 1 h after the
impact.

%Your text goes here. Use the \LaTeX\ automatism for cross-references
%as well as for your citations, see Sect.~\ref{sec:1}.

\begin{figure}
\centering
\includegraphics[height=8cm]{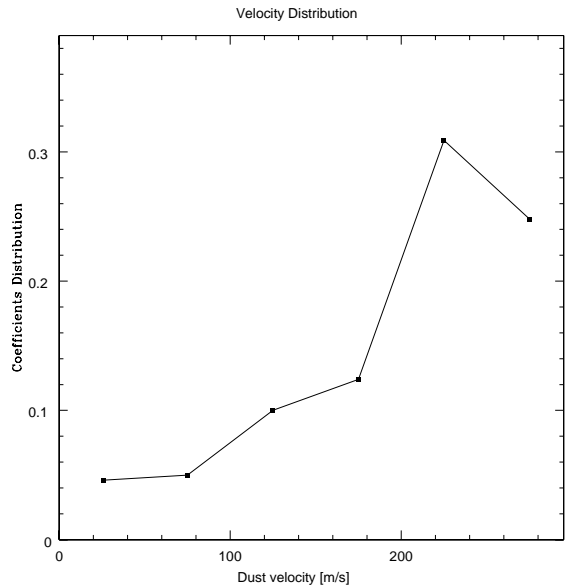}
%\picplace{5cm}{2cm} % Give the correct figure height and width in cm
%
\caption{Parameters c$_j$ versus dust terminal velocity. The
distribution peaks at 225 ms$^{-1}$.}
\label{fig:2}       % Give a unique label
\end{figure}

\section{Conclusion}

We developed a discrete linear inverse problem approach which allows
to simultaneously fit modeled light-curves to a set of observational
light-curves of cometary dust of 9P/Tempel 1 acquired by NAC aboard
Rosetta spacecraft. We derive a broad accurate velocity distribution
of the dust particles (between 1 and 600 ms$^{-1}$). It peaks at 225
ms$^{-1}$. This value is in good agreement with already published
estimates of the expansion velocity of the dust cloud. We consider
these successful comparisons as an evidence of the appropriateness
of the method.

%The large amount of material ejected from the comet~\cite{k05:ref}
The velocities of the dust seen in the light curve in the first
minutes after the impact suggest early acceleration of the ejected
material close to the cometary surface.

Despite the relative success of the models and approach presented in
this work, a few points may be raised that need further improvement.
In particular, the relative robustness of the method. Furthermore,
because of the unstable character of the ill-conditioned solutions,
further regularization algorithms for the stabilization of the
solution should be investigated. Limits on late acceleration (and
therefore in late production of material) also should be enquired,
work is in progress.

We acknowledge the funding of the national spaces agencies ASI,
CNES, DLR, the Spanish Space Program (Ministerio de Educacion y
Ciencia), SNSB, and ESA.

% For built-in environments use
%

%
% BibTeX users please use
\bibliographystyle{prsty}
\bibliography{ref}

\begin{thebibliography}{10}

\bibitem{a05:ref}
M.~F. {A'Hearn} {\it et~al.}, ``{Deep Impact: Excavating Comet Tempel 1},''
  Science {\bf 310,} 258--264 (2005).

\bibitem{k05:ref}
H.~U. {Keller} {\it et~al.}, ``{Deep Impact Observations by OSIRIS Onboard the
  Rosetta Spacecraft},'' Science {\bf 310,} 281--283 (2005).

\bibitem{2005Natur.437..987K:ref}
M. {K{\"u}ppers} {\it et~al.}, ``{A large dust/ice ratio in the nucleus of
  comet 9P/Tempel 1},'' Nature {\bf 437,} 987--990 (2005).

\bibitem{k06:ref}
H.~U. {Keller}, M. {K{\"u}ppers}, S. {Fornasier}, P.~J. {Gutierrez}, L.~J.
  Stubbe F.~{Hviid}, J. {Knollenberg}, S.~C. {Lowry}, M. {Rengel}, I.
  {Bertini}, and et~al., ``{Observations of Comet 9P/Tempel 1 around the Deep
  Impact event by the OSIRIS cameras onboard Rosetta},'' Icarus \ p.\ Available
  online 20 November 2006 (2006).

\bibitem{jorda06:ref}
L. {Jorda}, P. {Lamy}, G. {Faury}, H. {Keller}, S. {Hviid}, M. {K{\"u}ppers},
  D. {Koschny}, J. {Lecacheux}, P. {Gutierrez}, and L. {Lara}, ``{Properties of
  the dust cloud caused by the Deep Impact experiment},'' Icarus \ p.\
  Available online 17 November 2006 (2006).

\bibitem{LAPACK3:ref}
E. Anderson {\it et~al.}, {\em LAPACK User's Guide, Third Edition} (SIAM,
  1999), {\tt www.netlib.org/lapack}.

\bibitem{meech05:ref}
K.~J. {Meech} {\it et~al.}, ``{Deep Impact: Observations from a Worldwide
  Earth-Based Campaign},'' Science {\bf 310,} 265--269 (2005).

\bibitem{sugita05:ref}
S. Sugita {\it et~al.}, ``{Subaru Telescope Observations of Deep Impact},''
  Science {\bf 310,} 274--278 (2005).

\bibitem{schleicher06:ref}
D.~G. {Schleicher}, ``{Compositional and physical results for Rosetta's new
  target Comet 67P/Churyumov Gerasimenko from narrowband photometry and
  imaging},'' Icarus {\bf 181,} 442--457 (2006).

\bibitem{feldman06:ref}
P.~D. {Feldman}, S.~R. {McCandliss}, M. {Route}, H.~A. {Weaver}, M.~F.
  {A'Hearn}, M.~J. {Belton}, and K.~J. {Meech}, ``{Hubble Space Telescope
  observations of Comet 9P/Tempel 1 during the Deep Impact encounter},'' Icarus
  \ p.\ Available online 9 October 2006 (2006).

\end{thebibliography}
%
% Non-BibTeX users please follow the syntax
% the syntax of "referenc.tex" for your own citations
%\input{referenc}
%%%%%%%%%%%%%%%%%%%%%%%%%%%%%%%%%%%%%%%%%%%%%%%%%%%%%%%%%%%%%%%%%%%%%%  }

%%%%%%%%%%%%%%%%%%%%%%%%%%%%%%%%%%%%%%%%%%%%%%%%%%%%%%%%%%%%%%%%%%%%%%

\printindex
\end{document}